# B.SAR – Blind SAR Data Focusing
## *for low cost applications*

Cataldo Guaragnella
DEI – Department of Electric and Information Engineering
Politecnico di Bari
Bari, Italy
cataldo.guaragnella@poliba.it

Tiziana D'Orazio
ISSIA – Institute of Intelligent Systems for Automation
CNR – Italian National Research Council
Bari, Italy
dorazio@ba.issia.cnr.it

*Abstract*— Synthetic Aperture RADAR (SAR) is a radar imaging technique in which the relative motion of the sensor is used to synthesize a very long antenna and obtain high spatial resolution. The increasing interest of the scientific community to simplify SAR sensors and develop automatic system to quickly obtain a sufficiently good precision image is fostered by the will of developing low-cost/light-weight SAR systems to be carried by drones. Standard SAR raw data processing techniques assume uniform motion of the satellite (or aerial vehicle) and a fixed antenna beam pointing sideway orthogonally to the motion path, assumed rectilinear. In the same hypothesis, a novel blind data focusing technique is presented, able to obtain good quality images of the inspected area without the use of ancillary data information. Despite SAR data processing is a well established imaging technology that has become fundamental in several fields and applications, in this paper a novel approach has been used to exploit coherent illumination, demonstrating the possibility of extracting a large part of the ancillary data information from the raw data itself, to be used in the focusing procedure. Preliminary results – still prone to enhancements – are presented for ERS raw data focusing. The proposed Matlab software is distributed under the Noncommercial – Share Alike 4.0 – International Creative Common license by the authors.

*Keywords*— *Drone systems; UAV; SAR systems on drones; efficient focusing of SAR data; Inverse problem; radar theory; remote sensing; SAR data focusing; phase shifts; range migration; satellite trajectory; spatial resolution; synthetic aperture radar; Geometry; Geophysics; Satellites; Focusing; Ancillary Data; Singular Value Decomposition; Blind deconvolution; Numerical analysis, Signal Processing; Parameter estimation; Algorithm; Imaging; Phase estimation; Phase compensation; Azimuth; Computational modeling; Computer Vision; Image resolution; Radar imaging; Synthetic Aperture*

I. INTRODUCTION

Synthetic Aperture Radar ([1], [2], [3]) is a remote sensing system used to obtain high-resolution images of the reflectivity properties of the Earth surface at the frequency used in the emitted signal. The Synthetic Aperture Radar can acquire very high resolution images of the inspected area using high bandwidth of the transmitted coherent illumination signal by means of an accurate processing of the ground received returns. In a standard structure, the system is composed of a platform (i.e. airborne or satellite) using the same antenna both for the transmitting and receiving phases; the target scene is repeatedly illuminated with pulses of radio waves. The signals echoes are received in the band of the transmitted pulse along the satellite track at different positions, converted to an intermediate frequency channel and IQ sampled so that a baseband processing procedure can allow the synthetic aperture processing algorithm, obtaining the equivalent return of a very narrow antenna beam. Due to physical limitations, it is not possible to manufacture an antenna of long length and mount it on an airborne platform, and this is particularly true for small and cheap unmanned Aerial Vehicles (UAV) so that the narrow beam in the along track direction (Azimuth) is synthesized from data.

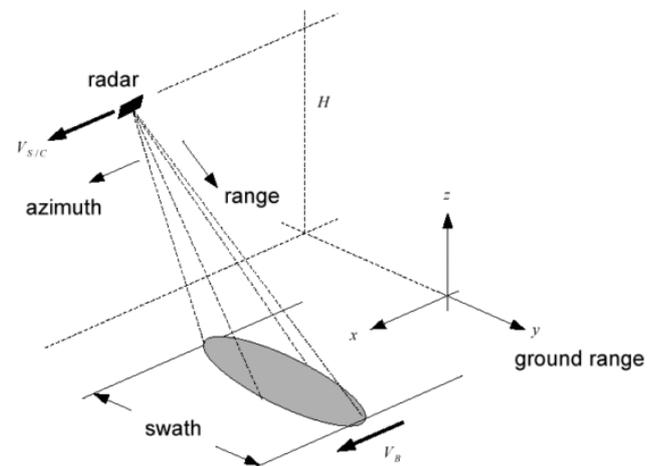

*Fig. 1: SAR data acquisition geometry*

Fig.1 reports the acquisition geometry of a SAR system. The synthesis procedure in focusing the acquired data is carried out by coherent integration. Each target on the ground contributes to the radar return on several subsequent transmitted pulses.

In SAR two main directions are important to focus the data: slant range direction, in which transmitted pulses travel, and Azimuth direction, i.e. the direction of sensor movement. The precise knowledge of the geometry of the acquisition allows to add in phase each contribute of the single point scatterer on the ground to obtain the focused image. The wider the beam, the less the detail acquired by any return, but the larger the integration size of the track contribution to synthesize the image, so that the azimuth radar resolution is not theoretically bounded. The practical azimuth resolution is limited by the PRF choice (the Pulse Repetition Frequency of transmitted pulses used for coherent illumination of the target area), i.e. the azimuth sampling frequency.

Modern hardware technology permits to reduce the dimensions and weights of SAR systems into small and cheap flying platforms that can be conveniently used also with low cost unmanned aerial vehicle (UAV) platforms and flying drones. High-resolution microwave images of the observed scene can be obtained under various environmental conditions. Thus, UAV-SAR attracts growing interest in recent years ([4], [5]).

The possibility of developing commercial low cost systems is anyway still limited by the complication of the development of SAR due to the precise need of mission parameters to obtain good quality images. Such parameters are very unstable for this kind of applications; furthermore the knowledge of all the mission parameters introduces a complication in the system in storage of ancillary parameters files, increasing their cost.

Despite the well known and precise algorithms like W-K ([7], [13]) and Range-Doppler (RD) ([8], [14]) for Synthetic Aperture Radar (SAR) have been developed and are commonly used by the space agencies all over the world (ESA, NASA, etc.), in this paper a proposal for a novel, blind algorithm for SAR data focusing is presented, able to obtain pretty good precision focusing and take care of Doppler Centroid (DC) shifts, requiring none of all the ancillary informations needed in precise SAR data focusing that make the generation of the SAR image always a tunable activity to be carried out by experts.

The main idea of the proposed approach is to exploit all the inherent information intrinsically stored in the data itself to extract the focusing reference functions to be used in a W-K or RD algorithm to obtain the Single Look Complex of any SAR sensor without even knowing important ancillary data parameters, needed by all the SAR focusing processors, such as the distance at the center of the beam, the radar sampling frequency, the transmitted chirp bandwidth, the chirp rate and the chirp duration, the radar wavelength, the PRF, the sensor speed and the off nadir angle, used in data acquisition. The proposed approach has been tested on several images of ERS raw data, made accessible for the scientific purpose from the Italian Space Agency (ASI) and the cross comparison with the state of the art focusing algorithm is carried out. Preliminary results seem to indicate a good accordance to the standard focusing of obtained images with respect to the officially distributed ones.

## II. SAR DATA ACQUISITION

SAR normally use chirp transmitted pulse. The radar moves along the track in the azimuth direction while transmitting repeated chirp signals with a given PRF in the direction orthogonal to the satellite trajectory (slant range). The received samples of the radar return are acquired along the slant range direction and each recorded data vector represents a row of a sequentially stored matrix of complex data (raw data matrix). The chirp signal is a linearly frequency modulated (FM) signal spanning the bandwidth of the radar; its peculiarity is the ability of conveying high energy of the transmitted long pulse duration without the limitation inherent in geometric resolution because the use of a matched filter at the receiver end allows to recover a high peak signal to noise ratios and a resolution in distance directly related to the chirp signal bandwidth ([1], [2], [3]). Due to the many returns contribution at each given instant of the recorded file (the superposition of returns of all the points on the ground within the physical antenna footprint located on the Earth at the same distance from the sensor), the acquired raw data matrix assumes the aspect of a complex random noise with a gaussian distribution.

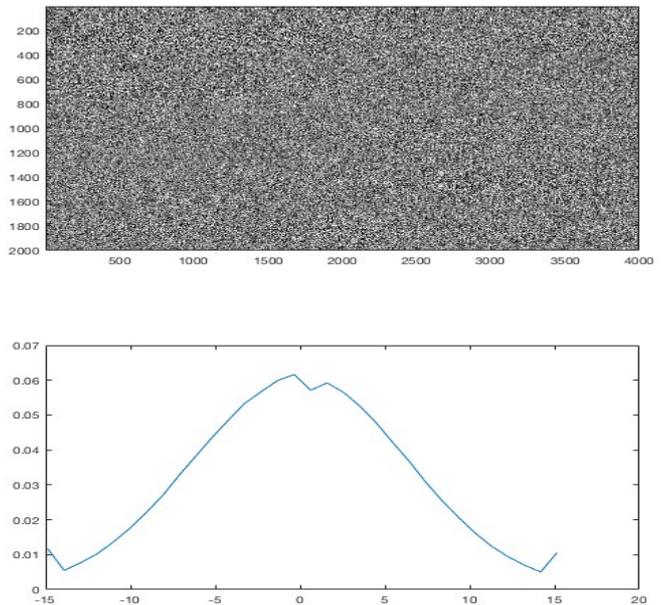

*Fig. 2: Raw data matrix (real part) and probability density function*

Fig.2 reports the real part of a raw data matrix and the probability density function of the acquired signal. The problem of SAR data focusing is commonly known as an inverse problem, where the deconvolution algorithm that permits to obtain the focused image makes recourse to the definition of the correlation operator to be used on raw data, precisely known if several mission parameters are known; such parameters are made available, for each RAW data product, in the form of an ancillary information file. As the SAR algorithm is linear, the deconvolution implementation can be described observing a single point target at the center of the scene (or within) and extract all the useful information from geometrical considerations.

### A. Transmitted Chirp

At every position along the sensor track of a perfectly absorbing surface where a single strong point scatterer is placed, a return of the scatterer to the chirp is recorded. This signal is acquired along the track in all the point until it remains within the azimuth antenna beam pattern. The antenna gain changes and assumes its largest value when the scatterer is exactly at the center of the antenna pattern while the intensity of the received return is tapered for different position of the sensor along the track. Fig. 3 reports the magnitude of the response of a simulated single point scatterer in an ideal acquisition SAR system. What is clearly visible in the figure is both the degradation of the response for azimuthal position displaced from the center beam and the phenomenon known as range migration or range walk, that is the displacement of the first sample of the radar return as the sensor moves along the track. In the figure is also represented the real part of the simulated raw data.

with respect to theoretical assumed ones have to be considered. A lot of refinement algorithms able to deal with such problems have been presented in the scientific literature and are part of the focusing algorithm actually available and used to produce official data ([9], [10], [11], [12]). Also efficiency has always been addressed in the performances evaluation of a focusing algorithm, nowadays overcome by the high computational capacity of existing hardware that make this problem less severe than once.

## III. EXTRACTING INFORMATION FROM SAR DATA

Several information about the SAR data format can be extracted from the acquired data. In this section all the possible information that can be extracted will be addressed and the required procedures to extract the information will also be presented.

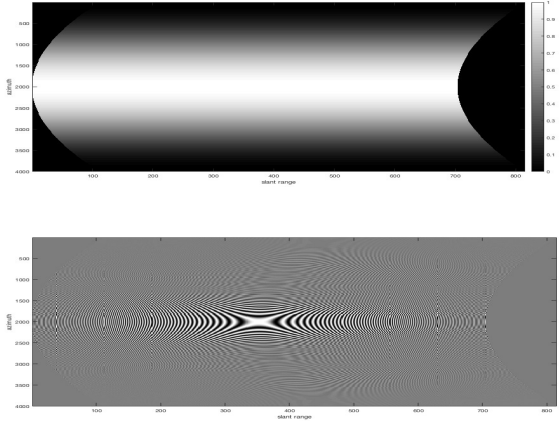

Fig. 3: Simulated response of a point scatterer: Magnitude and real part

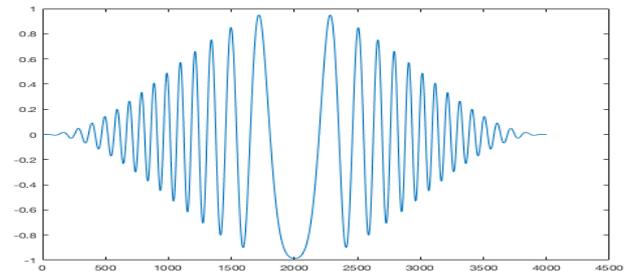

Fig. 4: Azimuth chirp, real part

### B. Range Migration and Azimuthal chirp

If the raw data matrix in Fig. 3 is read in the vertical (azimuth) direction, due to the range migration and delay in received chirp response of the single point scatterer on the ground, the azimuth response reveals to be a chirp signal too, amplitude modulated by the antenna beam pattern in azimuth (see Fig.4).

The state of the art focusing algorithms obtain high resolution images from SAR raw data matrices basing on precise knowledge of the data acquisition geometry and SAR system parameters. The mostly used SAR focusing algorithms are RD and W-K. Also a modified version of the focusing algorithm is the Chirp Scaling (CS). In the Range-Doppler approach, the focusing algorithm can be decomposed easily in two different parts: first focusing in the range direction is carried out by correlating the received raw data rows with the range reference function (i.e. the transmitted chirp signal); once obtained the range focused image, focusing in the azimuth direction takes place normally in two steps: first, the range migration compensation is applied (Range Cell Migration Compensation), then the azimuth compression takes place using a geometrically constructed phase history of the received signal.

In a real scenario, several other distortions can take place, due to non ideality of the acquisition geometry. To deal with problems like doppler centroid estimation and compensation, generally due to relative motion of Earth ground point with respect to the sensor trajectory or satellite attitude deviations

### A. Blind SAR data Focusing Algorithm

The acquired data, as discussed, is the result of backscattering contribution of the ground at the SAR frequency. As a coherent illuminating source is used, the received data refer to several observations of the same scene taken in different points along the satellite trajectory; each return should contain roughly the same information, so that the exploitation of coherence of the received signal can be attempted. The radar transmitted pulses are stable in time so that all the received returns show a strong azimuth correlation. This hypothesis allows to use some correlation-based algorithm to extract useful information from data. The use of Singular Value Decomposition (SVD) ([6]) technique can give us information about the reference functions to be used to focus the image.

### B. SVD – Signal Processing

Singular Value Decomposition, in its economy formulation, is a standard algorithm able to decompose a given rectangular matrix into the product of three matrices, U, S and V as schematically depicted in Fig. 5.

*X* is the data matrix of size $M \cdot N$, *U* and *V* are orthonormal matrices; *U* has the same size of the matrix *X* while *S* is a real valued diagonal matrix of size *N* and *V* is a complex orthonormal square matrix of size *N*, where:

$$U^H \cdot U = I_N$$

and  (1)

$$V^H \cdot V = I_N$$

**S** is the matrix containing the Singular Values of the matrix decomposition, sorted along the diagonal from the highest value to the lowest. In Analytical form, The SVD decomposition can be written simply as:

$$X = U \cdot S \cdot V^H \qquad (2)$$

where the superscript $(\cdot)^H$ represents the transpose and conjugate operator (Hilbert operator). If a right multiplication for matrix **V** of both terms in (2) is applied, it can be usefully restated in another form:

$$X \cdot V = U \cdot S = E \qquad (3)$$

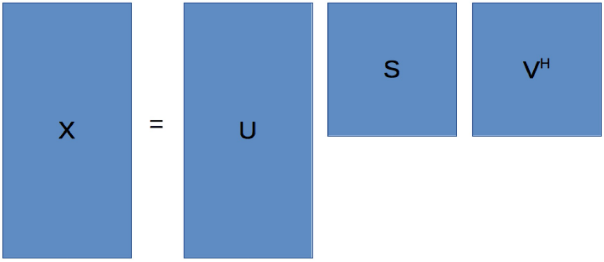

Fig. 5: Schematic description of a SVD matrix decomposition

The **V** matrix is the matrix mixing columns of the data matrix **X** to obtain **E**, an orthogonal matrix. The right multiplication of the matrix **U** with **S** only scales the vector columns in **U** non affecting the orthogonality property. **E** is a matrix containing columns obtained from the linear mixture of the columns of the data matrix. The columns of the original data matrix can be considered different realizations of a given process, so that this orthogonality and the hypothesis of zero mean for the signal time series permits the separation of the data matrix in uncorrelated signals. As a simple example if the SVD decomposition **X** of two columns requires the right singular vector matrix **V** an orthonormal matrix of size 2. The **V** matrix in this simple case represents a complex Gibbs rotation matrix. In this very simple case, the operation carried out by the decomposition becomes clear:

$$[\vec{x}_1, \vec{x}_2] \cdot \begin{bmatrix} c & -s \\ s^H & c \end{bmatrix} = [\vec{e}_1, \vec{e}_2] \qquad (4)$$

The **E** matrix is thus obtained as a simple linear combination of columns of **X**. **V** is the Gibbs rotation matrix and:

$$|c|^2 + |s|^2 = 1 \qquad (5)$$

In a geometrical representation we can consider *c* and *s* parameters able to scale, rotate and phase shift vectors they multiply so that their sum and difference mixtures gives the two orthogonal vectors in **E**.

The modula of such vectors represent the singular values and one can easily demonstrate that such values are proportional to the estimates of the standard deviation of the resulting signals in **E**. As in **E** the two transformed vectors are independent, the mostly correlated information in the two signals in **X** is transferred to the first singular vector in **E**, while the remaining part is in the second. When applied to a multicolumn matrix, this procedure tends to accumulate all the strongly correlated information on columns in the first left eigenvector, so that it contains the information in common on the image columns in the azimuth direction. The coherent illumination due to the transmission of the chirp produces very correlated information; in particular, in the azimuth direction the correlated information is the doppler phase history due to the scanning process. The expected result is that the first left singular vector should closely related to the doppler history of the SAR system.

Also, accordingly to the process of SVD, the right singular vectors contain the mixing coefficient able to orthogonalize the raw data matrix. This information is closely related to the transmitted information in the slant range direction. All the rows in the data matrix contain the same information, i.e. the transmitted chirp, delayed and phase shifted of an amount depending on the SAR geometry. As the data are strongly correlated, the SVD decomposition will store automatically in the rotation matrix **V** all the coefficients able to phase shift the subsequent transmitted chirps so that their sum convey the most important part of the global transmitted energy of the pulse.

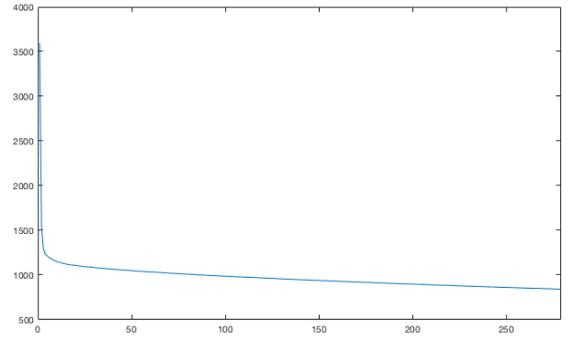

Fig. 6: Plot of the Singular Values for the SAR raw data matrix decomposition

The coefficient must then be closely related to both the geometrical properties of the SAR acquisition and the characteristics of the transmitted chirp. This consideration will be illustrated in the next paragraph. Basing on these observations, a simple and direct scheme to obtain a good focusing of the raw data in a blind mode has been devised.

## IV. EXPERIMENTAL RESULTS

The SVD decomposition is then applied as a test of the proposed approach to several ERS raw data matrices without any knowledge of the mission parameters. The plot of the diagonal elements of matrix S is a simple description of how the energy of the orthogonalized signals is distributed. This gives a quick description of how efficient was the separation of signals into uncorrelated components. Fig.6 shows the plot of singular values for the ERS data matrix.

A large part of the energy in the data matrix is concentrated in the first singular value, clearly stating that the first left singular vector (i.e. the first column of matrix **U**) should contain the orthogonal signal with maximum energy in the data. Fig.7 represents the real part of this singular vector and, as it can be seen, it is clearly similar to the azimuth chirp, hence showing also the antenna beam pattern intensity modulation. From the simple observation of the first left singular vector of the matrix $U$ two informations can be extracted: the antenna beam pattern in azimuth and the phase history of the azimuth reference to be used in the focusing procedure.

Thresholding can be applied to define the useful phase history and estimate the antenna beam pattern. Also, the estimate of the doppler centroid can be obtained measuring the azimuth position of the lowest instantaneous frequency with respect to the peak value of the azimuth reference function, representing the antenna beam pattern central position.

Considering the matrix product of the SVD decomposition,

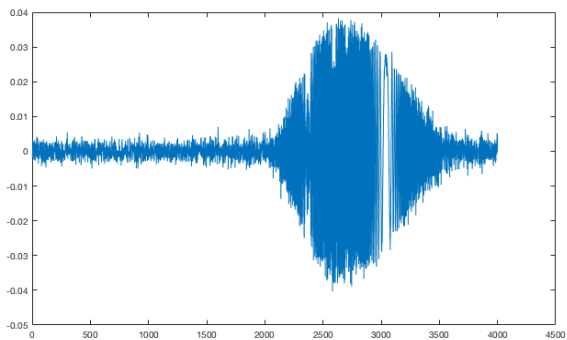

*Fig. 7: First left singular vector containing the Azimuth chirp extracted from raw data – real part*

to extract the first left singular vector of matrix $U$, the mixing coefficients to be used are the ones in the right first singular vector of matrix $V$.

This signal contains the complex coefficients needed to *rephase* and add the received chirp signal in the raw data matrix. It is interesting to verify what are such coefficients. Fig. 8 represents the real part of the first column of the matrix $V$.

The proposed approach is thus simple and direct and allows to extract useful information to focus the received data. It should be pointed out that the possibility of obtaining good estimates of the range and azimuth chirp responses are due to the clear presence of a point scatterer with high signal to noise ratio that conveys a large part of the data matrix energy in a restricted part of the data matrix, allowing good estimates. Once known parameters of the azimuth and range histories, focusing can become simple and can be carried out with either RD or W-K algorithms. In this paper the frequency approach has been used to obtain the focused image. In the subsequent subparagraph the detail of the proposed algorithm is given.

### A. B-SAR - Blind SAR Data focusing Algorithm

From the matrix $U$ two informations can be extracted: the antenna beam pattern in azimuth and the phase history of the azimuth reference to be used in the focusing procedure. Thresholding can also be applied to define the useful phase history and estimate the antenna beam pattern. Also, observing the azimuth position of the lowest instantaneous frequency with respect to the azimuth antenna beam center the estimate of the doppler centroid can be obtained. Fig. 7 and Fig. 8 show that the obtained information is noisy so that a more accurate derivation of the azimuth and range histories can be carried out to obtain a clean focused image.

To select the proper length of the range chirp history and reconstruct a clean reference in range a simple thresholding is used as the reference signal is much higher than the background noise. In this case the selected range time duration by the choice of a threshold at 10% of the signal peak value has revealed efficient. The estimated length of the range chirp was 703 samples. To reduce the influence of noise, the unwrapped phase of the range chirp has been used in the LMS estimation of the parabolic phase in the construction of the range chirp to be used in the focusing algorithm.

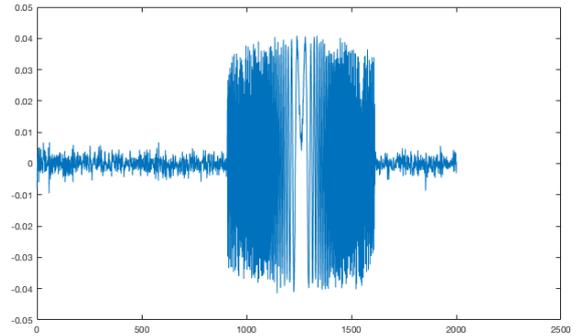

*Fig. 8: First right singular vector containing the Range chirp extracted from raw data – real part*

The same procedure was carried out for the azimuth chirp, with a slight more care: the antenna beam pattern estimated in this way is not always effective due to the growing attenuation and the joint influence of azimuth and range beam patterns with the slant range and the possible presence of extended strong scatterers that can reduce the quality of the estimated pattern. Also, it is not clear where the azimuth phase history should be stopped. The main objective is to limit the phase history in a proper way to avoid azimuth aliasing.

The proposed approach uses the phase unwrapping of the azimuth history. The cut of the important information was obtained also by thresholding with respect to the antenna beam pattern maximum, occurring at the signal peak position and selecting the phase history in a generally asymmetric interval around the peak. The limitation of this interval was chosen as the 10% of the estimated beam pattern. Also, tapered tails are used to avoid aliasing effects on the focused image, both in the range and azimuth reference functions.

Once obtained via LMS the references of azimuth and range histories (see Fig.9), we can consider ready to focus the image. In Fig.10, the focused ERS image obtained with the proposed algorithm (B-SAR) and the focusing obtained by a standard Range-Doppler SAR processor are compared.

A block diagram of the complete algorithm is reported in Fig. 12.

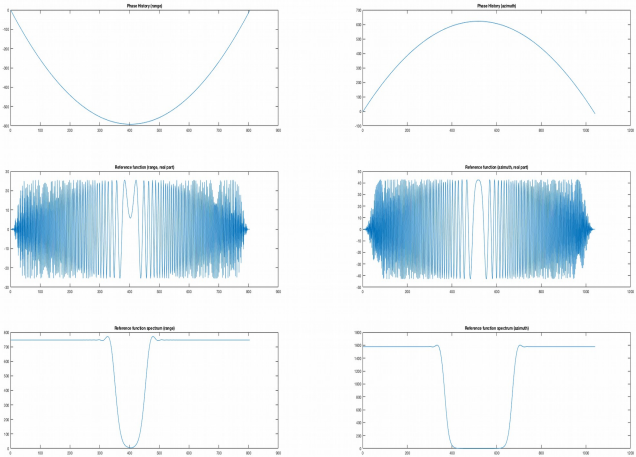

*Fig. 9: Range and Azimuth references to be used in the focusing procedure. Left range, right azimuth. The three rows present extrated phase histories, the real parts of the reference functions and their spectra*

To quantify the obtained results the extraction and oversampling of the response of the point scatterer in the scene has been produced.

## V. Discussion, limitations and future developments

In this work the SVD decomposition has been used to extract correlated information from SAR raw data on scenes where a strong point scatterei is present. The use of the SVD is sufficient to develop a simple and direct procedure to focus the acquired data without the need of precise information about the sensor attitudes, path and SAR systems parameters. The proposed algorithm at the state of the art is sufficient to obtain a fair focused image even if not a complete exploitation of the correlated data has been yet devised. This task will be the goal of future research. Also, the possibility of blind focusing SAR raw data, here addressed only in the presence of a point scatterer (e.g. a corner reflector or a transponder), in the general case of SAR stripmap data focusing will be addressed. The development of similar procedures also for SPOT and SCAN SAR data would be interesting and these problem will be addressed in the future.

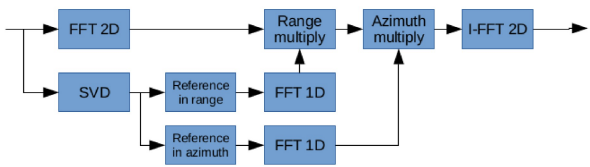

*Fig. 10: B.SAR Data Processing block diagram*

## VI. Conclusion

Fair maps can be obtained by focusing SAR stripmaps raw data with the proposed algorithm. Obtained results are close to the standard images given by the official space agencies, even if a thorough comparison and cross check is still to be carried out. The proposed algorithm, developed in Matlab, is distributed under the Noncommercial – Share Alike 4.0 – International Creative Common license by the authors. Tuning and complete data correlation exploitation will be carried out in a future work.

### Acknowledgment

The author wish to thank Dr. Fabio Bovenga of CNR-ISSIA, Bari (the Institute for the Study of Intelligent Systems for Automation of Italian National Research Council) and Dr. Raffaele Nutricato of GAP s.r.l. (Geophysical Applications and Processing, Bari) for the useful discussion on the method and for the ERS images used to test the algorithm, Dr. Luigi Dini of ASI (Italian Space Agency) for his support and useful suggestions.

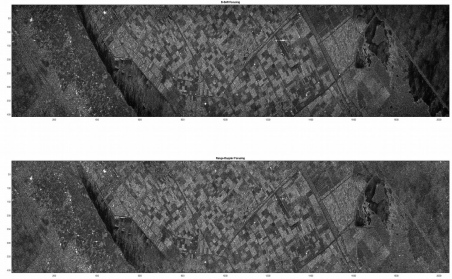

*Fig. 11: A sample of focused image with the proposed algorithm and the standard focusing obtained via Range Doppler algorithm. ERS 1 – Matera See also: https://www.facebook.com/SST.PoliBa/photos/a.532377260203317.1073741830.499905266783850/1552996644808035/?type=3&theater*

References

[1] Ridenour, L. N.: "Radar System Engineering," McGraw-Hill Book Co.,New York, 1945, or Boston Technical Lithographers, Lexington, Mass., 1963.

[2] Skolnik, M. I.: "Introduction to Radar Systems," McGraw-Hill Book Co., New York,1962.

[3] Klauder, J. R. et al.: "The Theory and Design of Chirp Radars," Bell System Technical Journal, July 1960.

[4] C. J. Li and H. Ling, High-resolution, downward-looking radar imaging using a small consumer drone, 2016 IEEE International Symposium on Antennas and Propagation (APSURSI)}, pp. 2037-2038, doi 10.1109/APS.2016.7696725

[5] Brian D. Rigling, Flying blind: a challenge problem for SAR imaging without navigational data, Proc. SPIE, vol. 8394, doi 10.1117/12.923439

[6] Golub, G. H. and Reinsch, C., Singular value decomposition and least squares solutions, Numerische Mathematik, 1970, Apr, 01, vol. 14, no.5, pp. 403-420, issn 0945-3245, doi 10.1007/BF02163027

[7] C. Cafforio and C. Prati and F. Rocca, IEEE Transactions on Aerospace and Electronic Systems, SAR data focusing using seismic migration techniques, 1991, vol. 27, no 2, pp. 194-207, doi 10.1109/7.78293, ISSN 0018-9251,

[8] R. Bamler, IEEE Transactions on Geoscience and Remote Sensing, A comparison of range-Doppler and wavenumber domain SAR focusing algorithms, 1992, vol 30,no 4, pp. 706-713, doi10.1109/36.158864, ISSN 0196-2892

[9] S. Amein and J. J. Soraghan, A new chirp scaling algorithm based on the fractional Fourier transform, IEEE Signal Processing Letters, 2005, vol 12,no 10, pp 705-708, doi 10.1109/LSP.2005.855547, ISSN 1070-9908

[10] A. S. Amein and J. J. Soraghan, Fractional Chirp Scaling Algorithm: Mathematical Model, IEEE Transactions on Signal Processing, 2007, vol 55, no 8, pp. 4162-4172, doi 10.1109/TSP.2007.895994, ISSN 1053-587X

[11] F. Li and S. Li and Y. Zhao, Focusing Azimuth-Invariant Bistatic SAR Data With Chirp Scaling, IEEE Geoscience and Remote Sensing Letters, 2008, vol 5, no 3, pp. 484-486, doi 10.1109/LGRS.2008.921743, ISSN 1545-598X